\pdfoutput=1

\documentclass[11pt]{article}

\usepackage[final]{acl}

\usepackage{times}
\usepackage{latexsym}

\usepackage[T1]{fontenc}

\usepackage[utf8]{inputenc}

\usepackage{microtype}

\usepackage{inconsolata}

\usepackage{graphicx}

\usepackage{amsmath, amsthm}
\usepackage{amssymb}
\usepackage{booktabs}
\usepackage{mathtools}
\usepackage{bbm}
\usepackage[utf8]{inputenc}
\usepackage[utf8]{inputenc}
\usepackage{algorithm}
\usepackage{algpseudocode}
\usepackage{xcolor}
\usepackage{hyperref}
\usepackage{multirow}
\usepackage{colortbl}

\newtheorem{theorem}{Theorem}
\newtheorem{lemma}[theorem]{Lemma}
\newtheorem{definition}{Definition}

\newcommand{\diff}[2]{%
  \raisebox{-0.4ex}{\footnotesize\textcolor{#1}{#2}}%
}
\usepackage[dvipsnames]{xcolor}

\usepackage{tcolorbox}
\tcbuselibrary{listings,breakable}

\usepackage{tabularx}
\usepackage{booktabs} 
\usepackage{array}    
\newcommand{\re}[1]{{\color{black}{#1}}}

\title{Differentially Private Synthetic Text Generation for Retrieval-Augmented Generation (RAG)}

\author{
 \textbf{Junki Mori\textsuperscript{1}},
 \textbf{Kazuya Kakizaki\textsuperscript{1}},
 \textbf{Taiki Miyagawa\textsuperscript{1}},
 \textbf{Jun Sakuma\textsuperscript{2,3}},
\\
 \textsuperscript{1}NEC Corporation,
 \textsuperscript{2}Institute of Science Tokyo,
\\
 \textsuperscript{3}RIKEN Center for Advanced Intelligence Project
\\
 \texttt{\{junki.mori,kazuya1210,miyagawataik\}@nec.com} 
 \\
 \texttt{sakuma@c.titech.ac.jp}
}

\begin{document}
\maketitle
\begin{abstract}
Retrieval-Augmented Generation (RAG) enhances large language models (LLMs) by grounding them in external knowledge. However, its application in sensitive domains is limited by privacy risks. 
Existing private RAG methods typically rely on query-time differential privacy (DP), which requires repeated noise injection and leads to accumulated privacy loss. 
To address this issue, we propose DP-SynRAG, a framework that uses LLMs to generate differentially private synthetic RAG databases. 
Unlike prior methods, the synthetic text can be reused once created, thereby avoiding repeated noise injection and additional privacy costs. 
To preserve essential information for downstream RAG tasks, DP-SynRAG extends private prediction, which instructs LLMs to generate text that mimics subsampled database records in a DP manner.
Experiments show that DP-SynRAG achieves superior performance to the state-of-the-art private RAG systems while maintaining a fixed privacy budget, offering a scalable solution for privacy-preserving RAG.

\end{abstract}

\section{Introduction}

\begin{figure}[t]
  \includegraphics[width=\columnwidth]{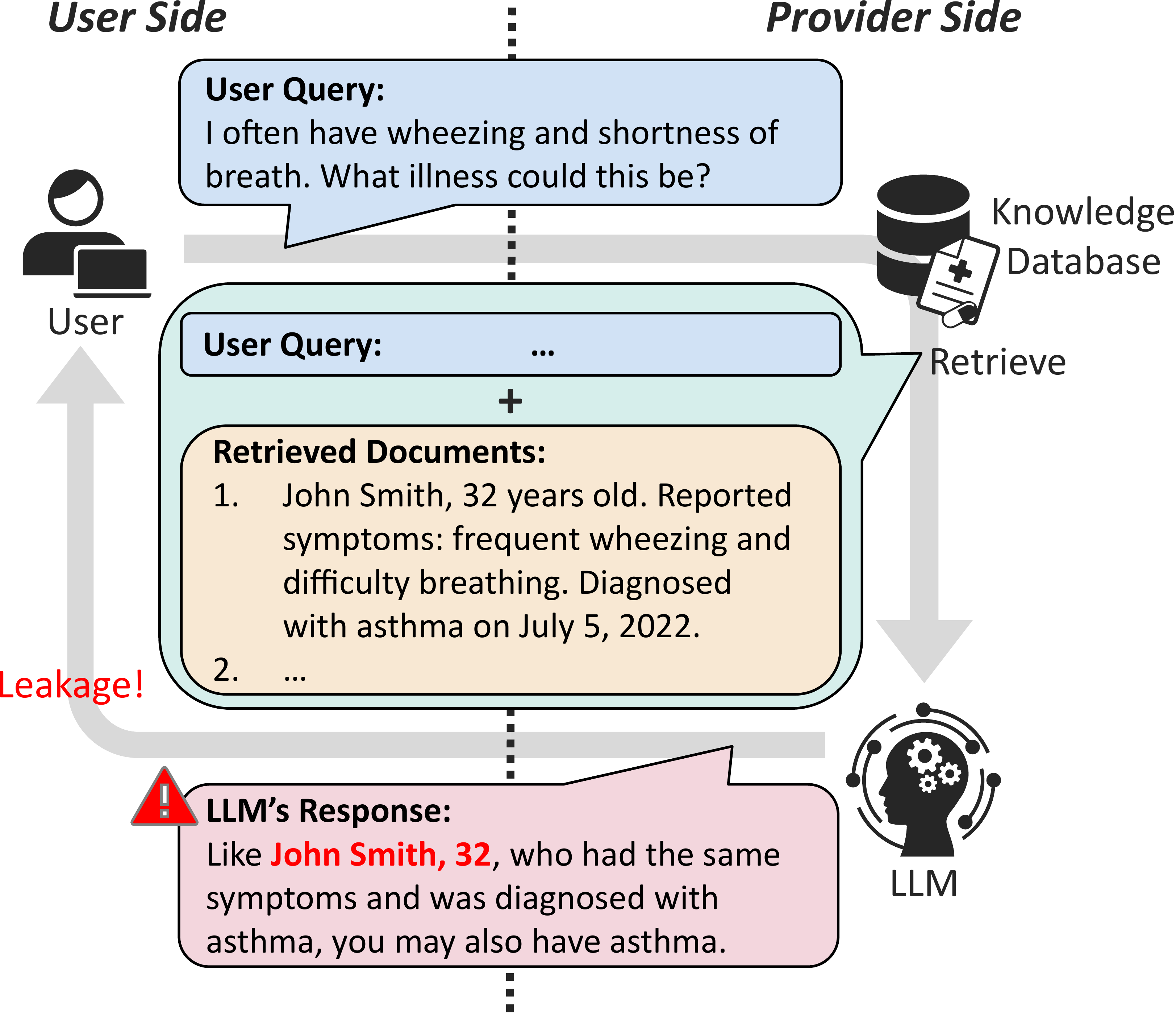}
  \caption{A demonstration of privacy risks in RAG databases: sensitive information contained in retrieved documents (e.g., patient names) may be revealed to benign users through LLM’s responses.}
  \label{fig:setting}
\end{figure}

Retrieval-Augmented Generation (RAG) \cite{rag} has been widely used to enhance the performance of large language models (LLMs) \cite{rag_survey} by leveraging external knowledge databases. However, recent studies have identified significant privacy risks in RAG systems when their databases contain sensitive information \cite{zeng24,qi25,jiang25,dimaio24,wang25}.
For instance, extraction attacks against medical chatbots for patients or recommender systems for customers can expose private data to attackers. Moreover, sensitive information in the retrieved documents may also be inadvertently revealed to benign users during normal interactions through LLM outputs (Figure~\ref{fig:setting}).

To develop secure RAG systems, differential privacy (DP), a formal framework for protecting individual records, has begun to be used \cite{grislain25,koga25,wang25dprag}.
Existing approaches ensures DP by injecting noise into the LLM’s responses to user queries, thereby reducing the influence of any single record.

However, these methods must add noise to each response; thus, they consume the privacy budget proportionally to the number of queries. Consequently, in typical RAG scenarios involving many queries under a fixed privacy budget, the utility of responses degrades substantially. 

To avoid repeated noise injection, we propose a text generation method using LLMs, called \textbf{D}ifferentially \textbf{P}rivate \textbf{Syn}thetic text generation for \textbf{RAG} databases (DP-SynRAG). Once synthetic texts with DP guarantees are generated, they can be reused indefinitely as the RAG database without incurring additional privacy budget, regardless of the number of queries. To achieve high-quality private text generation, we adopt \textit{private prediction} \cite{hong24,tang24,amin24,gao25}, which prompts the LLM with subsampled database records and rephrasing instructions, while perturbing the aggregated output token distributions to limit per-record information leakage. One limitation of these approaches is that they capture only global properties of the entire dataset, discarding not only sensitive details but also important knowledge needed for RAG. 
DP-SynRAG mitigates this issue by clustering documents in a DP manner based on keywords and document-level embeddings, thereby grouping semantically similar documents and separating distinct topics. Applying private prediction to these clusters in parallel enables 
large-scale synthetic text generation that preserves cluster-specific knowledge at low privacy cost.
Finally, to reduce the effect of low-quality synthetic texts, we apply LLM-based self-filtering to improve downstream RAG performance.

We validate our approach on three datasets tailored to our setting. Results show that our method outperforms existing private RAG approaches in most cases while maintaining a fixed privacy budget, demonstrating its effectiveness and scalability for privacy-preserving RAG applications.

\section{Related Works}
\subsection{Privacy Risks of RAG}
When sensitive information is stored in the external databases used by RAG, various privacy risks arise. The first major threat is that adversarial prompts can trigger extraction attacks, causing personally identifiable information (PII) or raw database text to be leaked through the LLM’s output to malicious users \cite{zeng24,qi25,jiang25,dimaio24,peng25}. 
\citet{wang25} has also shown that extraction may occur even from benign queries.
The second major threat is membership inference attacks (MIAs), in which an adversary infers whether specific target data exist in the database. Several studies \cite{li25,anderson25,naseh25,liu25} have shown that such attacks are also effective against RAG systems.

Our paper proposes a general defense method that counters all the above attacks by ensuring DP.

\subsection{Privacy-preserving RAG}
Various methods have been proposed to protect the privacy of RAG. One early approach paraphrases documents in the database using LLMs to remove sensitive information \cite{zeng25mitigatin}. However, this method ignores privacy risks during retrieval and lacks DP guarantees. As non-DP defenses, researchers have proposed training embedding models robust to adversarial queries \cite{he25}, detecting MIA queries and excluding the corresponding documents \cite{choi25}, and perturbing data embeddings \cite{yao25}.

Recent studies have also proposed directly enforcing DP on LLM outputs \cite{grislain25,koga25,wang25dprag}. However, because each query output is perturbed, these methods consume the privacy budget per query, which increases the risk of leakage as queries accumulate.

\subsection{Differentially Private Text Generation}
Our method relies on DP-guaranteed synthetic text generation. Generated data can be used for downstream tasks such as training, in-context learning, and RAG without incurring extra privacy costs.

Early approaches apply local differential privacy (LDP) sanitization to raw text \cite{yue21,chen23}, but this severely reduces utility due to the strong noise to each word.

Recent work instead leverages the generative capabilities of LLMs to produce DP synthetic text. One direction is \textit{private fine-tuning}, where an LLM fine-tuned on private data using DP-SGD generates synthetic text under DP guarantees \cite{yue23,kurakin24,yu24}. These methods, however, are computationally expensive due to DP-SGD and impractical for RAG, where the underlying knowledge database is periodically updated, making repeated retraining infeasible.

Another direction is \textit{private prediction}, which applies a DP mechanism to the output token distribution when an LLM paraphrases the original text. This is typically implemented via \textit{subsample-and-aggregation} \cite{nissim07}, where the private dataset is divided into disjoint subsets, and non-private predictions from each subset are privately aggregated. Private prediction has been applied to generate a small number of texts for prompt tuning or in-context learning \cite{hong24,tang24,gao25}, while \cite{amin24} proposes generating large-scale data without assuming a specific downstream task. 
\re{In contrast to token-level private prediction, Aug-PE \cite{aug-pe} selects LLM-generated texts whose embeddings are close to those of the private data, ensuring DP at the selection stage.}
However, these methods capture only average dataset properties, which is insufficient for RAG applications. 

Accurate query answering requires synthetic data that preserves locality, i.e., the distinctive features of the original dataset. Current private prediction methods lack mechanisms for generating locality-aware outputs, leaving a significant gap in their applicability to RAG. Our method fills this gap. 
Although a recent study \cite{amin2025clusteringmedianaggregationimprove} reports that clustering improves synthetic data quality, their approach assumes public cluster centers, which is unrealistic in a private RAG setting as considered in this work.

\section{Preliminaries}
\subsection{Differential Privacy}
\label{sec:dp}
Let $\mathcal{D}$ be a set of possible datasets. Two datasets $D$ and $D'$ are called \textit{neighbors} if one is obtained from the other by dropping a single record. 

\begin{definition}[$(\varepsilon,\delta)$-DP]
A randomized mechanism $\mathcal{M}$ is $(\varepsilon,\delta)$-differentially private if any neghboring dataset $D,D'\in\mathcal{D}$ and any set $\mathcal{S}$ of possible outputs, it holds that
\[
\mathrm{Pr}[\mathcal{M}(D)\in\mathcal{S}]\leq e^{\varepsilon}\mathrm{Pr}[\mathcal{M}(D')\in\mathcal{S}]+\delta.
\]
\end{definition}




In this paper, we employ two representative DP algorithms: the \textit{Gaussian mechanism} and the \textit{exponential mechanism}.
The Gaussian mechanism adds Gaussian noise \( \mathcal{N}(0, \sigma^2 I_d) \) to a real-valued function \( f(D)\in\mathbb{R}^d \). The exponential mechanism selects an output $y$ with probability proportional to \( \exp\left(\frac{\varepsilon \cdot u(D, y)}{2\Delta_{\infty} u}\right) \), where \( u(D, y) \) is the utility function and $\Delta_{\infty} u = \sup_{y,D, D'} |u(D,y) - u(D',y)|$ denotes its sensitivity. Both mechanisms satisfy $(\varepsilon,\delta)$-DP (see Appendix~\ref{app:analysis}).





\subsection{Retrieval-Augmented Generation}
RAG combines information retrieval with LLMs to improve the factual accuracy of responses. It guides generation by conditioning the LLM on relevant information retrieved from an external corpus.

Formally, let $D=\{d_i\}_{i=1}^N$ denote a document corpus, and let $\mathcal{V}$ be a vocabulary space. Given a user query $q \in \mathcal{V}^*$, RAG first uses an embedding model $\mathcal{E}:\mathcal{V}^* \to \mathbb{R}^d$ to encode both the query and each document $d_i \in D$ into a shared vector space. A similarity function $\operatorname{sim}(\mathcal{E}(q), \mathcal{E}(d_i))$ ranks the documents, enabling the selection of the top-$k$ most relevant contexts. The retrieved documents $(d_{i_1}, \dots, d_{i_k})$ are concatenated with the query to form an augmented prompt:
$p_{\rm{aug}} = (q, d_{i_1}, \dots, d_{i_k})$.
The LLM $\mathcal{L}$ then conditions its generation on $p_{\rm{aug}}$ and repeatedly draws tokens from the token space $\mathcal{T}$ according to
\begin{align}
    \label{sampling}
    y_n \sim \operatorname{softmax}(\mathcal{L}(p_{\rm{aug}}, y_{<n})/\tau),
\end{align}
where $y_{<n}$ is the sequence of previously generated tokens and $\tau \geq 0$ is the temperature. 
The function $\mathcal{L}$ maps input tokens to a logit vector in $\mathbb{R}^{|\mathcal{T}|}$.

\subsection{Private Prediction}
\label{sec:private_prediction}



To generate DP synthetic data, our method relies on the private prediction framework, which enables LLMs to produce DP outputs. It has been applied to inference tasks that protect training data \cite{flemings24} and RAG databases \cite{koga25,grislain25,wang25dprag}, and DP synthetic data generation that preserves the privacy of original datasets \cite{amin24,hong24,tang24,gao25}.

This framework follows a subsample-and-aggregate paradigm: a randomly selected subset of the private dataset is divided into disjoint partitions, from which non-private predictions are obtained and then privately aggregated. We leverage the inherent uncertainty of token sampling in LLMs to perform private aggregation without adding explicit noise, thereby reducing output distortion \cite{amin24}. The key insight is that sampling tokens from the aggregation of clipped token logits via softmax can be viewed as an exponential mechanism. Let $D_s \subset D$ be a subsampled subset. For each $d_i \in D_s$, a prompt $p_i = (p, d_i)$ is constructed using a task-specific prompt template $p$. The logit of the $n$-th token is computed for each $p_i$, clipped to the range $[ -c, c ]$, and summed:
\[
z_n(D_s) = \sum_{d_i\in D_s} \operatorname{clip}_c \left(\mathcal{L}(p_i,y_{<n})\right).
\]
Sampling a token from $z_n(D_s)$ via softmax, as described in Eq.~(\ref{sampling}), constitutes an exponential mechanism with sensitivity $\Delta_{\infty}z_n=c$. Thus, the sequence $y_{\leq T}$ generated by repeatedly applying this process satisfies DP \cite{amin24}.

\section{Proposed Method}
Existing approaches to our problem \cite{grislain25,koga25} respond to user queries in RAG systems via private prediction mechanisms. However, when directly applied to multiple queries, their total privacy budget grows linearly with the number of queries. To address this, we propose DP-SynRAG, which generates synthetic data resembling the private data in advance through private prediction while ensuring DP. Since using this synthetic data as a knowledge corpus for later RAG inference is post-processing, the privacy budget remains fixed regardless of the number of queries.


Current methods for generating synthetic data via private prediction often produce tokens that capture only the average characteristics of randomly subsampled subsets of the original data (see Section \ref{sec:private_prediction}). While this supports domain-specific data generation, it often loses the fine-grained factual details useful as RAG context. In contrast, DP-SynRAG first clusters the dataset based on keywords and document-level embeddings under DP, grouping semantically similar documents and separating distinct topics. Applying private prediction to these subsets in parallel can generate large volumes of high-quality synthetic texts covering diverse topics in the original dataset at a low privacy budget.




\subsection{Problem Formulation}

We consider the problem of returning a privacy-preserving response to a user query $q$ under the RAG framework, where retrieval dataset $D=\{d_i\}_{i=1}^N$ is private. Formally, our goal is to design a mechanism that outputs differentially private sequence of $T$ tokens $y_{\leq T}$ with respect to $D$, while satisfying the two objectives simultaneously:

\noindent  \textbf{1. Privacy-budget efficiency:} 
The mechanism should satisfy DP with a fixed privacy budget independent of the number of queries, ensuring scalability in multi-query RAG scenarios.

\noindent  \textbf{2. RAG-specific utility:}  
The generated responses should exhibit high utility in terms of fact-based effectiveness in downstream RAG tasks rather than generic language-model metrics (e.g., perplexity).

    

\subsection{Overview of DP-SynRAG}
\begin{figure*}[t]
  \centering
  \includegraphics[width=\linewidth]{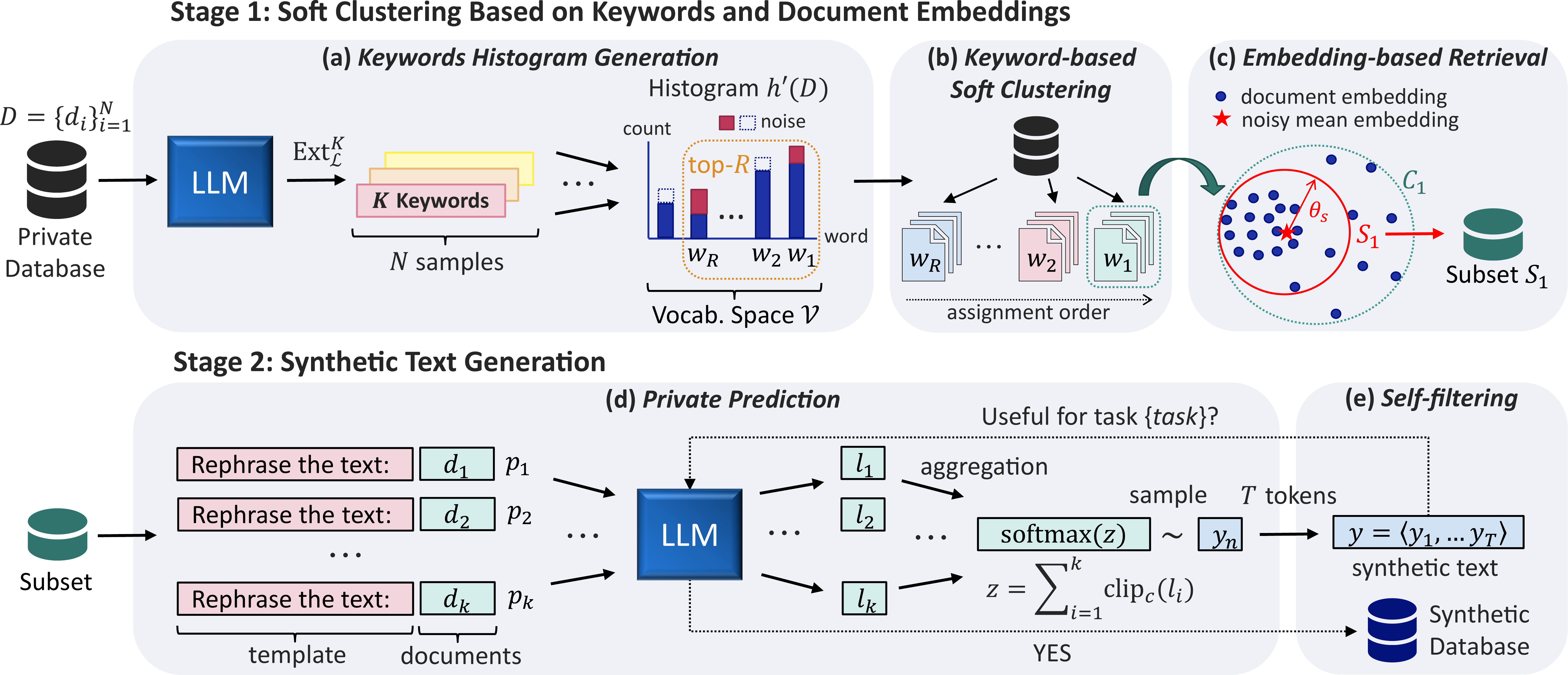}
  \caption{A two stage pipeline of \textbf{DP-SynRAG}. Stage 1 first constructs a noisy histogram from the $K$ keywords extracted from each document (a). Each document is assigned to up to $L$ clusters formed by the top-$R$ keywords from the histogram (b). From these clusters, relevant subsets are retrieved using embeddings (c). Stage 2 generates DP synthetic text by rephrasing the documents in each subset and privately aggregating the clipped output token logits (d). Finally, the LLM filters the synthetic texts based on their usefulness for the downstream task (e).}
  \label{fig:method}
\end{figure*}

Figure~\ref{fig:method} illustrates an overview of DP-SynRAG. The core idea is to partition the dataset into coherent topical subsets and apply rephrasing within each subset with DP guarantees.
Specifically it comprises two differentially private stages: (1) soft clustering based on keywords and document embeddings and (2) synthetic text generation. The full algorithm is shown in Appndix~\ref{app:algorithm} (Algorithm~\ref{algorithm}).

Stage 1 identifies representative keywords under DP (Figure~\ref{fig:method} (a)) and uses them to softly cluster documents into multiple topic-specific subsets (Figure~\ref{fig:method} (b)). To ensure each subset contains semantically similar documents, we further refine them with embedding-based similarity (Figure~\ref{fig:method} (c)). 

Stage 2 leverages LLMs to generate synthetic texts for each subset in parallel (Figure~\ref{fig:method} (d)). By rephrasing documents in a DP manner, we create synthetic data that retains the semantic richness of the original corpus while protecting sensitive information. A post-processing self-filtering step further improves the utility of the synthetic dataset for downstream RAG inference (Figure~\ref{fig:method} (e)).

\subsection{Stage 1: Soft Clustering Based on 
Keywords and Document Embeddings} 


\paragraph{(a) Keywords Histogram Generation.} 
This step privately constructs a histogram to extract representative keywords for forming clusters.
From each document, we first extract a set of $K$ distinct keywords that best represent the document instead of counting all words. 
This extraction from the vocabulary space $\mathcal{V}$ is performed by prompting the LLM $\mathcal{L}$ to select $K$ representative keywords, and formulated as the function $\operatorname{Ext}_{\mathcal{L}}^K \colon \mathcal{V}^* \rightarrow \{0,1\}^{|\mathcal{V}|}$, where $\sum_{v \in \mathcal{V}} \operatorname{Ext}_{\mathcal{L}}^K(d_i)_v = K$, indicating the $K$ keywords extracted from document $d_i$.
Summing across all documents yields a histogram $h(D) = \sum_{d_i \in D} \operatorname{Ext}_{\mathcal{L}}^K(d_i)$. To release $h(D)$ under DP, we add Gaussian noise to $h(D)$:
\[
h'(D) = h(D) + \mathcal{N}(0, \sigma_h^2 I_{|\mathcal{V}|}).
\]

\paragraph{(b) Keywords-based Soft Clustering.} 
From the histogram $h'(D)$, we select the top-$R$ most frequent keywords, denoted by $W = \{w_1, \dots, w_R\}$, where $w_ 1$ is the most frequent and $w_R$ the least frequent among $W$.
Each keyword $w_r$ defines a cluster $C_r$, and a document is assigned to $C_r$ if and only if it contains $w_r$, subject to the constraint that each document may belong to at most $L$ clusters. To reduce the dominance of uninformative high-frequency words, cluster assignment proceeds in reverse order of frequency, from $w_R$ up to $w_1$. Formally, for keyword $w_r$ ($r=R,\dots,1$), we define
\[
C_r = \Big\{ d_i \in D \Big\vert\ w_r \in d_i, \sum_{r'>r}\mathbbm{1}[d_i \in C_{r'}] <L \Big\},
\]
where $\mathbbm{1}[\cdot]$ is the indicator function. \re{This keyword-based design is practical under DP: once representative keywords are selected from the noisy histogram, cluster assignment is deterministic and incurs no additional privacy costs.}

In this way, a document may belong to at most $L$ clusters anchored by relatively infrequent but representative keywords. 
\re{In practice, a document may contain multiple informative keywords, and forcing it into a single cluster can associate it with a coarse or only partially relevant topic. Allowing up to $L$ overlapping assignments increases the likelihood that the document contributes to the subset that best matches its salient content, rather than being assigned according to a high-frequency but less informative keyword.}

\paragraph{(c) Embedding-based Retrieval.}
By keywords-based clustering, documents with similar topics are grouped together. However, some documents remain as outliers at the document-level, thereby introducing noise in subsequent synthetic data generation. To mitigate this, we remove outliers from each cluster in parallel based on document-level similarity. First, we compute the mean embedding $\mu(C_r)$ of each cluster via the Gaussian mechanism:
\[
\mu(C_r) = \sum_{d_i\in C_r} \mathcal{E}(d_i) + \mathcal{N}(0, \sigma_\mu^2I_d),
\]
where $\mathcal{E}(d_i)$ is the normalized embedding of $d_i$. This mean embedding\footnote{We omit division by cluster size to ensure $\mu(C_r)$ is defined even when the cluster is empty, and to keep the noise level consistent across clusters. A constant multiplicative factor does not affect similarity computations.} reflects the dominant characteristics of the cluster. We then privately retrieve the top-$k$ documents most similar to this embedding. 
The similarity threshold $\theta_s \in [0,1]$ is selected using the exponential mechanism with privacy parameter $\varepsilon_{\theta_s}$, as in \cite{grislain25}.
The utility function with sensitivity $1$ is defined as
\begin{equation}
    \label{threshold}
    u(C_r, \theta_s) = -\Big|\sum_i\mathbbm{1}\left[\theta_s\in[0,s_{ir}]\right]-k\Big|,
\end{equation}
where $s_{ir}=\operatorname{sim}(\mathcal{E}(d_i), \mu(C_r))$ is the similarity between $\mathcal{E}(d_i)$ and $\mu(C_r)$. Using the selected threshold, we retrieve the relevant subset $S_r \subseteq C_r$ as
\[
S_r = \{ d_i \in C_r \mid \operatorname{sim}(\mathcal{E}(d_i), \mu(C_r)) > \theta_{s} \}.
\]

\subsection{Stage 2: Synthetic Text Generation.}
\paragraph{(d) Private Prediction.}
By applying the private prediction method from Section~\ref{sec:private_prediction} in parallel to each subset using the LLM $\mathcal{L}$ and a rephrasing prompt template $p$ (e.g., ``Rephrase the following text:''), we generate a total of $R$ synthetic texts. Specifically, for each subset $S_r$, clipped and summed logits for the $n$-th token are computed as $z_n(S_r) = \sum_{d_i\in S_r} \operatorname{clip}_c \left(\mathcal{L}(p_i,y_{r,<n})\right)$, where $p_i = (p, d_i)$. These logits are used to sequentially generate a token sequence $y_{r,\leq T}$ of length $T$ via the exponential mechanism.



For the clipping method, we follow the approach of \citet{grislain25}, which emphasizes tokens with larger logit values and thereby reduces the impact of noise.
The details are given in Appendix~\ref{app:private_pred}.



\paragraph{(e) Self-filtering.} 
Synthetic text generated from small subsets is often low-quality, as tokens are sampled from a nearly random distribution, which increases the likelihood that useful information for downstream RAG tasks is lost. Because such text introduces noise, we apply self-filtering using LLMs. In methods like Self-RAG \cite{self-rag}, unrelated documents are removed after retrieval based on their query relevance; however, this approach increases inference-time computational cost. Therefore, we instead prompt the LLM with non-private downstream task information and the synthetic text, then filter the text according to whether it contains information essential for solving the downstream task prior to inference. The filtered outputs are then used to construct the synthetic RAG database. This filtering serves as a post-processing step in synthetic text generation.

\section{Privacy Analysis}
\label{sec:analysis}


Our algorithm generates $R$ synthetic texts with minimal total privacy budget, using the overlapping parallel composition introduced by \citet{smith21} and converted to zCDP (Appendix~\ref{app:analysis}). The formal privacy guarantee is stated below. We sketch the proof here and defer the full version to Appendix~\ref{app:analysis}.
\begin{theorem}
\label{thrm}
DP-SynRAG (Algorithm~\ref{algorithm} in Appendix~\ref{app:algorithm}) satisfies $(\varepsilon,\delta)$-DP for any $\delta > 0$ and $\varepsilon = \rho + \sqrt{4 \rho \log(1/\delta)}$, where
\[
\rho = \frac{K}{2\sigma_h^2} + L\left(\frac{1}{8}\varepsilon_{\theta_s}^2 + \frac{1}{2\sigma_{\mu}^2} + \frac{T}{2}\left(\frac{c}{\tau}\right)^2\right).
\]

\begin{proof}[Proof Overview.]
We adopt zero-concentrated differential privacy (zCDP) \cite{bun16}, a variant of DP, as it provides tighter composition bounds and more precise privacy accounting.
Our algorithm comprises two sequentially composed mechanisms: (a) histogram generation ($M_{\text{hist}}$) and (b-d) keyword-based clustering followed by parallel operations on each cluster ($M_{\text{clus}}$). Within $M_{\text{clus}}$, each cluster undergoes a sequence of operations: (c) retrieval ($M_{\text{retr}}$), (d) private prediction ($M_{\text{pred}}$). We exclude self-filtering from the privacy analysis since it a post-processing. Our proof proceeds in three steps.
First, we show that $M_{\text{hist}}$, $M_{\text{retr}}$, and $M_{\text{pred}}$ each satisfy $\rho_{\text{hist}}$, $\rho_{\text{retr}}$, and $\rho_{\text{pred}}$-zCDP, respectively, since they rely on Gaussian or exponential mechanisms (or their compositions). 
Next, each algorithm executed in parallel within a cluster satisfies $(\rho_{\text{retr}}+\rho_{\text{pred}})$-zCDP by sequential composition, and $M_{\text{clus}}$ satisfies $L(\rho_{\text{retr}}+\rho_{\text{pred}})$-zCDP due to overlapping parallel composition with $L$ overlaps.
Finally, by sequentially composing $M_{\text{hist}}$ and $M_{\text{clus}}$, the entire algorithm satisfies $\rho$-zCDP with $\rho = \rho_{\text{hist}} + L(\rho_{\text{retr}} + \rho_{\text{pred}})$. We then convert $\rho$-zCDP to $(\varepsilon,\delta)$-DP using the conversion lemma \cite{bun16}.

\end{proof}

\end{theorem}

\section{Experiments}

\begin{table*}[th]
\centering
\scalebox{0.8}{
\begin{tabular}{l l c c c c}
\hline
\multirow{2}{*}{\centering \textbf{Dataset}} & \multirow{2}{*}{\centering \textbf{Method}} & \multirow{2}{*}{\centering \textbf{Privacy Budget}} & \multicolumn{3}{c}{\textbf{Model}} \\
\cline{4-6}
& & & \textbf{Phi-4-mini} & \textbf{Gemma-2-2B} & \textbf{Llama-3.1-8B} \\
\hline
\multirow{6}{*}{\textbf{Medical Synth}}
& Non-RAG    & $\varepsilon_{\text{total}} = 0$      & 0.00\,\diff{black}{0.00}     & 0.00\,\diff{black}{ 0.00}     & 0.00\,\diff{black}{ 0.00}     \\
& RAG        & $\varepsilon_{\text{total}} = \infty$ & 87.00\,\diff{black}{ 0.00} & 85.20\,\diff{black}{ 0.00} & 86.20\,\diff{black}{ 0.00} \\
& DP-Synth \cite{amin24}   & $\varepsilon_{\text{total}} = 10$     & 0.00\,\diff{black}{ 0.00}     & 0.00\,\diff{black}{ 0.00}     & 0.00\,\diff{black}{ 0.00}     \\
& \re{Aug-PE \cite{aug-pe}}   & \re{$\varepsilon_{\text{total}} = 10$}     & \re{0.00}\,\diff{black}{ 0.00}     & \re{0.00}\,\diff{black}{ 0.00}     & \re{0.00}\,\diff{black}{ 0.00}     \\
& \textbf{DP-SynRAG (Ours)}       & $\varepsilon_{\text{total}} = 10$     & 67.26\,\diff{black}{ 2.22} & 67.06\,\diff{black}{ 1.68}  & 61.26\,\diff{black}{ 2.33}  \\
\cline{2-6}
& \multirow{2}{*}{DP-RAG \cite{grislain25}} & $\varepsilon_{\text{query}} = 10$ & \multirow{2}{*}{59.92\,\diff{black}{ 0.44}} & \multirow{2}{*}{67.06\,\diff{black}{ 0.44}} & \multirow{2}{*}{48.94\,\diff{black}{ 0.38}} \\
& & ($\varepsilon_{\text{total}} \approx 10000$) & & & \\
\hline
\multirow{6}{*}{\textbf{Movielens}}
& Non-RAG    & $\varepsilon_{\text{total}} = 0$      & 22.60\,\diff{black}{ 0.00}     & 34.00\,\diff{black}{ 0.00}     & 43.60\,\diff{black}{ 0.00}     \\
& RAG        & $\varepsilon_{\text{total}} = \infty$ & 67.80\,\diff{black}{ 0.00} & 54.60\,\diff{black}{ 0.00} & 70.80\,\diff{black}{ 0.00} \\
& DP-Synth \cite{amin24}  & $\varepsilon_{\text{total}} = 10$     & 37.60\,\diff{black}{ 2.60}     & 16.64\,\diff{black}{ 2.29}     & 46.12\,\diff{black}{ 2.54}     \\
& \re{Aug-PE \cite{aug-pe}}  & \re{$\varepsilon_{\text{total}} = 10$}     & \re{36.16}\,\diff{black}{ 3.79}     & \re{26.04}\,\diff{black}{ 5.64}     & \re{44.96}\,\diff{black}{ 2.10}     \\
& \textbf{DP-SynRAG (Ours)}       & $\varepsilon_{\text{total}} = 10$     & 42.56\,\diff{black}{ 1.97}  & 41.08\,\diff{black}{ 2.19}  & 54.12\,\diff{black}{ 2.51}  \\
\cline{2-6}
& \multirow{2}{*}{DP-RAG  \cite{grislain25}}   & $\varepsilon_{\text{query}} = 10$     & \multirow{2}{*}{34.72\,\diff{black}{ 0.54}} & \multirow{2}{*}{40.48\,\diff{black}{ 0.48}} & \multirow{2}{*}{56.80\,\diff{black}{ 0.62}} \\
& & ($\varepsilon_{\text{total}} \approx 5000$) & & & \\
\hline
\multirow{5}{*}{\textbf{SearchQA}}
& Non-RAG    & $\varepsilon_{\text{total}} = 0$      & 65.69\,\diff{black}{ 0.00}     & 70.59\,\diff{black}{ 0.00}     & 88.24\,\diff{black}{ 0.00}     \\
& RAG        & $\varepsilon_{\text{total}} = \infty$ & 92.16\,\diff{black}{ 0.00} & 94.12\,\diff{black}{ 0.00} & 95.10\,\diff{black}{ 0.00} \\
& DP-Synth \cite{amin24}  & $\varepsilon_{\text{total}} = 10$     & 60.20\,\diff{black}{ 2.91}     & 20.20\,\diff{black}{ 3.15}     & 40.00\,\diff{black}{ 3.88}     \\
& \textbf{DP-SynRAG (Ours)}       & $\varepsilon_{\text{total}} = 10$     & 89.61\,\diff{black}{ 3.22} & 85.10\,\diff{black}{ 2.13} & 91.18\,\diff{black}{ 3.25} \\
\cline{2-6}
& \multirow{2}{*}{DP-RAG \cite{grislain25}}   & $\varepsilon_{\text{query}} = 10$     & \multirow{2}{*}{85.10\,\diff{black}{ 1.75}} & \multirow{2}{*}{83.14\,\diff{black}{ 1.75}} & \multirow{2}{*}{84.90\,\diff{black}{ 1.78}} \\
& & ($\varepsilon_{\text{total}} \approx 1000$) & & & \\
\hline
\end{tabular}
}

\caption{Performance comparison across datasets, methods, and models under fixed total privacy budgets $\varepsilon_{\rm{total}}$, except for DP-RAG, which uses per-query budget $\varepsilon_{\rm{query}}$. The number of queries is 1,000 for Medical Synth, 500 for Movielens, and 102 for SearchQA. We report mean and standard deviation of the accuracy (\%) over 5 runs.}
\label{tab:privacy_comparison}
\end{table*}

\subsection{Settings}
\textbf{Datasets.} 
We evaluate our method on three datasets (details in Appendix~\ref{app:experiment_dataset}). \textbf{Medical Synth} \cite{grislain25} is a synthetic medical records dataset containing 100 fictitious diseases. It includes patient queries describing symptoms and corresponding doctor responses. Doctor responses to other patients serve as the private knowledge base. Performance is measured by accuracy, defined as whether the LLM’s output includes the correct fictitious disease name based on retrieved diagnoses from prior patients.
\textbf{Movielens-1M} \cite{movielens} is used in a natural language form to study privacy in RAG, as it includes user profiles as well as movie ratings. Using GPT-5, we generate textual descriptions of each user’s preferences from their profiles and favorite movies, defined as their top-10 rated movies. Private RAG documents include each user’s profile, generated preferences, and liked movies. The task is to recommend movies for a query user by referring to favorites of similar users. For simplicity, we restrict the dataset to the 30 most frequently rated movies. Accuracy is measured by whether the LLM’s output includes any of the user’s top-10 favorites.
\textbf{SearchQA} \cite{searchqa}, a standard RAG benchmark, consists of Jeopardy!-derived question–answer pairs with associated search snippets. We use training questions with at least 40 supporting snippets containing the correct answer, grouped into six bins by snippet count (40–50 to 90–100). we randomly sample 17 questions from each bin, yielding 102 in total.

\paragraph{Compared Methods.} 
We compare our \textbf{DP-SynRAG} with four approaches, each illustrating a different privacy–utility trade-off:
(1) \textbf{Non-RAG} excludes any RAG database ($\varepsilon=0$) and relies solely on the LLM’s general knowledge, representing the lower bound of utility.
(2) non-private \textbf{RAG} uses RAG database without privacy constraints ($\varepsilon=\infty$), representing the upper bound of utility.
(3) \textbf{DP-RAG} \cite{grislain25} is a representative private RAG approach that operates under a fixed privacy budget, which accumulates over multiple queries.
(4) \textbf{DP-Synth} \cite{amin24} is a representative DP-based synthetic data generation method that operates under a fixed privacy budget independent of the number of queries and belongs to the same category as DP-SynRAG.
\re{(5) \textbf{Aug-PE} \cite{aug-pe} is a DP synthetic text generation baseline in the same category as DP-Synth, but it does not rely on token-level private prediction during decoding. We evaluate Aug-PE on Medical Synth and MovieLens, but exclude it from SearchQA because Aug-PE assumes topic-focused generation, whereas SearchQA spans diverse open-domain topics.}

\paragraph{Models.} As an embedding model, we use multi-qa-mpnet-base-dot-v1 model (109M parameters) from the Sentence-Transformers library \cite{sentence-bert}, designed for semantic search. We compare three LLMs for text generation: Phi-4-mini-instruct
(3.8B) \cite{phi4mini}, Gemma-2 (2B) \cite{gemma2}, Llama-3.1 (8B) \cite{llama3}.

\paragraph{Implementation Details.}
Unless otherwise specified, the overall privacy budget is fixed at $\varepsilon_{\rm{total}} = 10$. For DP-RAG, we consider a per-query budget $\varepsilon_{\rm{query}}$; if the number of queries is $m$, then $\varepsilon_{\rm{total}} \approx m \varepsilon_{\rm{query}}$. We set  $\delta = 10^{-3}$ for all datasets. For RAG, the number of retrieved documents is $k=10$ for Medical Synth and SearchQA, and $k=15$ for Movielens. The inference temperature is fixed at $0$. The hyperparameters of the proposed and baseline methods are tuned on the validation set (see Appendix~\ref{app:_experiment_hyp} for details). \re{Our experimental pipeline is based on the public DP-RAG implementation,\footnote{\url{https://github.com/sarus-tech/dp-rag}} which we use for the DP-RAG baseline and as a common framework for implementing the other methods.} Each method is executed five times, and the average result is reported.
We build the public vocabulary space using the NLTK (v3.9.1) English word corpus \cite{nltk}, excluding common stopwords.

\begin{figure*}[t]
  \centering
  \includegraphics[width=\linewidth]{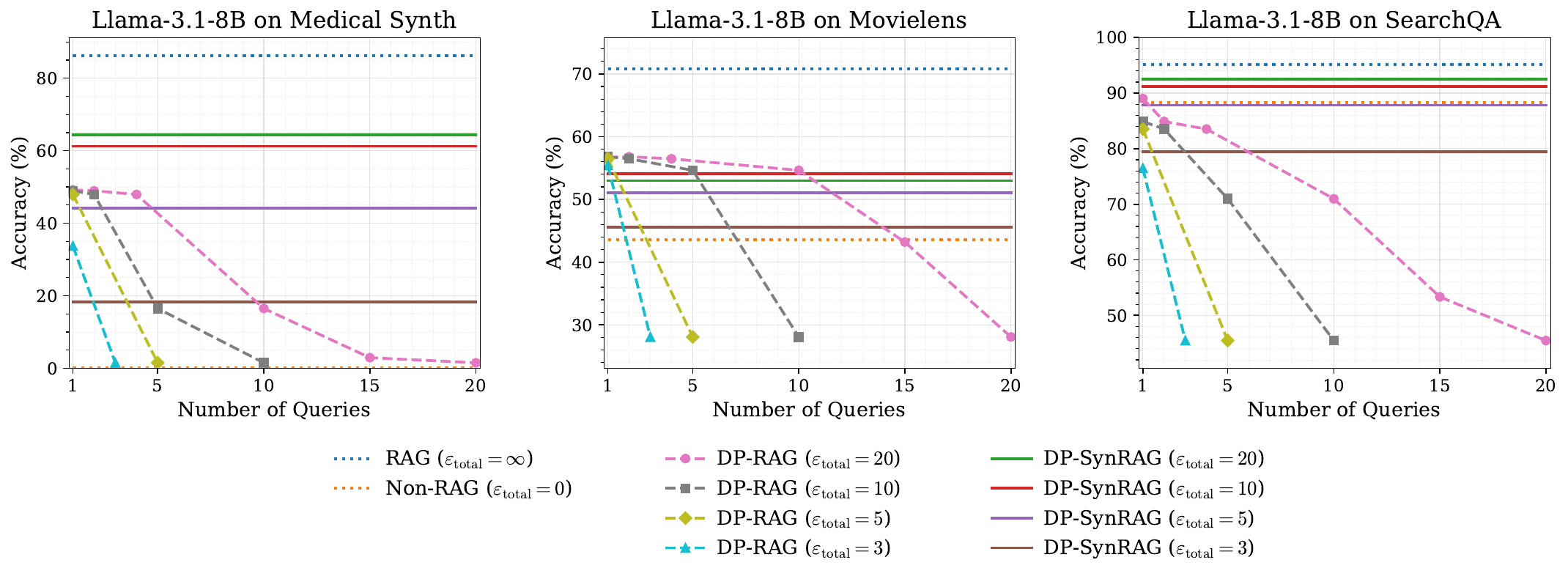}
  \caption{Accuracy versus number of queries under various fixed total privacy budgets. Since DP-SynRAG can reuse generated synthetic data as a RAG database without incurring additional privacy costs, its accuracy remains constant regardless of the number of queries. In contrast, DP-RAG needs to allocate a smaller privacy budget per query as the number of queries increases, causing its accuracy to decrease significantly. }
  \label{fig:accuracy_vs_query}
\end{figure*}

\subsection{Results}
\label{sec:results}
\paragraph{Main Results.}
Table~\ref{tab:privacy_comparison} presents the average accuracy across three datasets and three models. DP-SynRAG substantially outperforms Non-RAG by exploiting the RAG database while ensuring DP. In particular, Medical Synth requires answers containing fictitious disease names, which standalone LLMs lacking domain knowledge cannot handle, while DP-SynRAG achieves over 60\% accuracy across all models. As expected, compared with DP-Synth \re{and Aug-PE}, which generate synthetic texts that reflect only average characteristics and are therefore unsuitable for RAG, DP-SynRAG effectively retains critical information (e.g., disease names) in the database. Furthermore, even when the total privacy budget of DP-SynRAG equals the per-query budget of DP-RAG, DP-SynRAG demonstrates superior performance in most cases. Note that the total budget of DP-RAG far exceeds $\varepsilon_{\rm{total}} = 10$, as its total budget scales with the number of queries.

\re{We additionally report direct quality metrics of the synthetic texts in Appendix~\ref{app:synthetic_quality}. Although our primary objective is downstream RAG utility rather than generic language-model metrics, the perplexity results suggest that the synthetic texts generated by DP-SynRAG are of reasonable quality. Appendix~\ref{app:query_time_baselines} also compares DP-SynRAG with another query-time DP baseline.}

\paragraph{Accuracy vs. Number of Queries.}
To highlight a key advantage of DP-SynRAG, namely its constant privacy budget regardless of query count, we compare its accuracy with DP-RAG across different numbers of queries under fixed total privacy budgets ($\varepsilon_{\rm total}$). Figure~\ref{fig:accuracy_vs_query} presents results for three datasets using Llama-3.1. Each line depicts the accuracy of each method under different fixed $\varepsilon_{\rm{total}}$ values. When inference involves only a single query, both methods achieve comparable accuracy. However, as the number of queries increases, DP-RAG’s performance steadily declines; even at $\varepsilon_{\rm{total}}=20$, it fails completely once the query count reaches 20. These results demonstrate that generating synthetic text with DP guarantees is an effective strategy for RAG.

\paragraph{Impact of Dataset Redundancy.}
\re{To examine the effect of dataset redundancy, we additionally analyze Phi-4-mini on Medical Synth. For each query, we count the number of database documents containing the ground-truth disease name and group queries into bins accordingly. Table~\ref{tab:redundancy} reports the accuracy of DP-RAG and DP-SynRAG in each bin.}

\re{The accuracy of both methods improves as topic redundancy increases, but both methods fail on rare diseases supported by at most 30 documents. This trend suggests that dependence on redundancy is not specific to our clustering design, but rather reflects an inherent property of private RAG under DP, where the influence of each individual record must be bounded. From a privacy perspective, the low utility on rare diseases is expected: if a system could reliably answer such queries under DP, it would implicitly leak the existence of rare patient records, increasing re-identification risk. We therefore view reduced utility on rare topics as a reasonable and inherent privacy--utility trade-off in private RAG.}

\begin{table}[t]
\centering
\resizebox{\columnwidth}{!}{%
\begin{tabular}{lcccccc}
\toprule
Method & 0--30 & 30--60 & 60--90 & 90--120 & 120--150 & 150+ \\
\midrule
DP-RAG    & 0.00 & 13.73 & 41.83 & 62.55 & 81.18 & 75.06 \\
DP-SynRAG & 0.71 & 25.78 & 64.17 & 72.36 & 75.88 & 79.08 \\
\bottomrule
\end{tabular}
}
\caption{\re{Accuracy (\%) on Medical Synth for Phi-4-mini, grouped by the number of database documents that contain the ground-truth disease name. Larger counts indicate greater topic redundancy in the private database.}}
\label{tab:redundancy}
\end{table}

\paragraph{Privacy of RAG.} 
We evaluate the privacy risks of RAG and the effectiveness of DP using Medical Synth to quantify instances in which sensitive information is leaked from the LLM outputs. Sensitive information is defined as the full names of patients contained in the database records. Specifically, we measure how often a patient’s full name appears in the model’s responses. 
The evaluation uses 1,000 benign queries employed in the performance evaluation and 100 targeted attack queries explicitly crafted to extract patient names. The targeted attack uses the prompt template: \textit{“Who is the patient diagnosed with \{disease\}?”} for 100 diseases.

Table~\ref{tab:leakage} reports the average number of leaks across five experimental runs for each method. The results indicate that even benign queries can cause RAG to inadvertently reveal patient names, demonstrating a clear privacy risk. Moreover, targeted attacks substantially increase the number of leaks. 
While DP-RAG significantly reduces leakage, a small number of leaks still occur under benign queries. In contrast, our proposed method achieves a sufficiently low probability of leakage. Even with $\varepsilon_{\rm{total}}=10$, the per-token privacy budget for synthetic text is sufficiently small to effectively prevent the disclosure of sensitive information.



\begin{table}[t]
\centering
\resizebox{\columnwidth}{!}{%
\begin{tabular}{l cc cc cc}
\hline
\multirow{2}{*}{\centering Method} 
& \multicolumn{2}{c}{Phi-4-mini} 
& \multicolumn{2}{c}{Gemma-2-2B} 
& \multicolumn{2}{c}{Llama-3.1-8B} \\
\cline{2-7}
& Benign & Attack 
& Benign & Attack 
& Benign & Attack \\
\hline
RAG & 4 & 85 & 12 & 90 & 22 & 81 \\
DP-RAG & 2.8 & 0 & 1.8 & 2 & 6 & 0 \\
DP-SynRAG & 0.5 & 1.25 & 0 & 0.25 & 2.4 & 1 \\
\hline
\end{tabular}
}
\caption{Average occurrences of patient full-name leakage in Medical Synth under 1,000 benign queries and 100 attack queries. We use $\varepsilon_{\rm{total}}=10$ for DP-SynRAG and $\varepsilon_{\rm{query}}=10$ for DP-RAG.}
\label{tab:leakage}
\end{table}


\begin{table}[t]
\centering
\resizebox{\columnwidth}{!}{%
\begin{tabular}{l c c c}
\hline
Method & Phi-4-mini & Gemma-2-2B & Llama-3.1-8B \\
\hline
\multicolumn{4}{c}{\textbf{Medical Synth}} \\
\hline
DP-SynRAG & 67.26 & 67.06 & 61.26  \\
w/o Retrieval & 65.92\diff{red}{↓1.34} & 61.46\diff{red}{↓5.60} & 57.74\diff{red}{↓3.52}  \\
w/o Self-filtering & 66.78\diff{red}{↓0.48} & 66.74\diff{red}{↓0.32} & 52.20\diff{red}{↓9.06}  \\
Hard clustering ($L=1$) & 42.52\diff{red}{↓24.74} & 51.40\diff{red}{↓15.66} & 29.38\diff{red}{↓31.88}  \\
\hline
\multicolumn{4}{c}{\textbf{Movielens}} \\
\hline
DP-SynRAG & 42.56 & 41.08 & 54.12  \\
w/o Retrieval & 42.28\diff{red}{↓0.28} & 42.84\diff{ForestGreen}{↑1.76} & 53.76\diff{red}{↓0.36}  \\
w/o Self-filtering & 40.68\diff{red}{↓1.88} & 38.40\diff{red}{↓2.68} & 45.12\diff{red}{↓9.00}  \\
Hard clustering ($L=1$) & 38.36\diff{red}{↓4.20} & 38.32\diff{red}{↓2.76} & 46.56\diff{red}{↓7.56}  \\
\hline
\multicolumn{4}{c}{\textbf{SearchQA}} \\
\hline
DP-SynRAG & 89.61 & 85.10 & 91.18  \\
w/o Retrieval & 89.22\diff{red}{↓0.39} & 83.73\diff{red}{↓1.37} & 90.98\diff{red}{↓0.20}  \\
w/o Self-filtering & 89.61 & 85.10 & 91.18  \\
Hard clustering ($L=1$) & 76.67\diff{red}{↓12.94} & 67.06\diff{red}{↓18.04} & 82.94\diff{red}{↓8.24}  \\
\hline
\end{tabular}
}
\caption{The average accuracy when each component of DP-SynRAG is removed: embedding-based retrieval, self-filtering, and soft clustering. The subscript indicates the accuracy difference between the full DP-SynRAG and the DP-SynRAG without each component.}
\label{tab:ablation}
\end{table}

\paragraph{Ablation Study.} 
The proposed method centers on keyword-based clustering and private prediction, operating independently of other components. To assess the impact of additional features, we perform an ablation study on three elements: document-based retrieval, soft versus hard clustering, and self-filtering. Table~\ref{tab:ablation} reports the accuracy when each element is disabled. The results indicate that these components enhance performance on most datasets. In particular, using hard clustering causes many documents to be grouped under irrelevant keywords, substantially degrading the quality of the generated synthetic text. Note that self-filtering is not applied to SearchQA because this dataset includes diverse question types rather than fixed tasks.

\section{Conclusion}
This study introduces DP-SynRAG, a novel framework for generating privacy-preserving synthetic texts for RAG that preserves both data utility and formal DP guarantees. By creating synthetic RAG databases, DP-SynRAG eliminates the need for repeated noise injection and enables unlimited query access within a fixed privacy budget. Experiments on multiple datasets show that DP-SynRAG consistently achieves performance better than existing private RAG methods in most cases, demonstrating its scalability and practical effectiveness.

\section*{Limitations}

\re{
While DP-SynRAG shows strong performance and scalability for privacy-preserving RAG, several limitations remain. First, the method is less effective when the RAG database contains only a few documents supporting a topic. As shown in Section~\ref{sec:results}, both DP-RAG and DP-SynRAG improve as topic redundancy increases, while both fail on rare diseases supported by at most 30 documents. This reflects an inherent property of private RAG under DP, where each record's influence must be bounded. In addition, our clustering may be less effective when semantically related documents share few surface words. More semantic-aware variants, such as expanding extracted keywords with synonyms or generating abstract topic descriptors with an LLM, may alleviate this issue while fitting naturally into our pipeline.


Second, like other methods based on token-level private prediction, DP-SynRAG experiences substantial utility loss under extremely tight privacy budgets, because privacy is enforced at the token level during synthetic text generation. At the same time, our comparison with Aug-PE suggests that avoiding token-level DP alone does not make a method suitable for RAG, since the non-token-level DP text generation methods fail to preserve task-critical private content. Developing non-token-level DP methods that remain effective for RAG is therefore an important direction for future work.

Third, our approach includes several hyperparameters that control clustering and noise calibration. However, as demonstrated in Appendix~\ref{app:_experiment_hyp}, fixed default values perform consistently well across different models and datasets, minimizing the need for extensive tuning.

Finally, when the RAG database is updated, the synthetic database must be refreshed to maintain privacy guarantees. This introduces additional preprocessing and generation cost, which can be important in frequently updated deployments. However, unlike private fine-tuning approaches, our method requires no model retraining, and the refresh can be limited to affected subsets of the database. We provide runtime and database-refresh details in Appendix~\ref{app:runtime_refresh}.}


\bibliography{custom}

\newpage
\appendix

\section{Algorithm Details}
\label{app:algorithm}
This section describes the details of our DP-SynRAG.
Algorithm~\ref{algorithm} shows the complete procedure. In the following, we explain in detail the algorithm components that require further clarification.

\begin{algorithm*}[pht]
\caption{Differentially Private Synthetic Text Generation for RAG (DP-SynRAG)}
\label{algorithm}
\begin{algorithmic}[1]
\State \textbf{Input:} Private database $D = \{d_i\}_{i=1}^N$, Vocabulary $\mathcal{V}$, Embedding model $\mathcal{E}$, LLM $\mathcal{L}$, \texttt{task}
\State \textbf{Parameters:} $K, J, L, \sigma_h, \sigma_\mu, \varepsilon_{\theta_s}, c, \tau, T, \theta_p$
\State \textbf{Output:} Synthetic database $D_{\mathrm{synth}}=\{y_j\}$ for RAG

\Statex \textcolor{blue}{\# Stage 1: (a) Keywords Histogram Generation}
\State $\operatorname{Ext}_{\mathcal{L}}^K(d_i) \in \mathbb{R}^{|\mathcal{V}|}, \ i=1,\dots,N$ \qquad \textcolor{gray}{$\triangleright$ Extract distinct $K$ keywords from $d_i$ instructing $\mathcal{L}$}
\label{alg:keyword}
\State $h(D) \leftarrow \sum_{d_i \in D} \operatorname{Ext}_{\mathcal{L}}^K(d_i)$ \qquad \textcolor{gray}{$\triangleright$ Histogram of keywords}
\State $h'(D) \leftarrow h(D) + \mathcal{N}(0, \sigma_h^2 I_{|\mathcal{V}|})$ 
\Statex \textcolor{blue}{\# Stage 1: (b) Keywords-based Soft Clustering}
\State $W = \{ w_1, \ldots, w_R \} 
   \leftarrow$ top-$R$ most frequent keywords from $h'(D)$ \quad \textcolor{gray}{$\triangleright$ Descending order}
\State $C_r \leftarrow \big\{ d_i \in D \big\vert\ w_r \in d_i, \ \sum_{r'>r}\mathbbm{1}[d_i \in C_{r'}] <L \big\},\ r=R,R-1,\dots,1$ \quad \textcolor{gray}{$\triangleright$ Define $R$ soft clusters}
\ForAll{cluster $C_r$}
    \Statex \quad \ \  \textcolor{blue}{\# Stage 1: (c) Embedding-based Retrieval}
    \State $\mu(C_r) \leftarrow \sum_{d_i\in C_r} \mathcal{E}(d_i) + \mathcal{N}(0, \sigma_\mu^2 I)$ \quad \textcolor{gray}{$\triangleright$ Compute mean embeddings}
    \State Select $\theta_{s}$ via exponential mechanism with utility function defined as:
    \Statex {\centering $u(C_r, \theta_s) = -\Big|\sum_i\mathbbm{1}\Big[\theta_s\in[0,\operatorname{sim}(\mathcal{E}(d_i), \mu(C_r))]\Big]-k\Big|$ \par}
    \State $S_r \leftarrow \{ d_i \in C_r \mid \operatorname{sim}(\mathcal{E}(d_i), \mu(C_r)) > \theta_{s} \}$ \quad \textcolor{gray}{$\triangleright$ Retrieve relevant documents}
    \Statex \quad \ \ \textcolor{blue}{\# Stage 2: (d) Private Prediction}
    \For{$n = 1$ to $T$}
        \State $p_i \leftarrow \texttt{concat}(p, d_i),\ d_i\in S_r$ \qquad \textcolor{gray}{$\triangleright$ Prompt: $p=$``Rephrase the following document:''}
        \label{alg:pred_start}
        \State $z_n(S_r) \leftarrow \sum_{d_i\in C_r} \operatorname{clip}_c \left(\mathcal{L}(p_i,y_{r,<n})\right)$ \qquad \textcolor{gray}{$\triangleright$ Compute clipped logits and sum them}
        \label{alg:pred_agg}
        \State $y_{r,n} \sim \operatorname{softmax} 
        (z_n(S_r)/\tau)$ \qquad \textcolor{gray}{$\triangleright$ Sample tokens}
        \label{alg:pred_end}
    \EndFor
    \State Set $y_r = (y_{r,1}, \ldots, y_{r,T})$
\EndFor

\Statex \textcolor{blue}{\# Stage 2: (e) Self-filtering}
\State $D_{\mathrm{synth}} \leftarrow []$
\label{alg:filter_start}
\ForAll{$y_r$ in $\{y_r\}_{r=1}^R$}
    \State $p_r \leftarrow \texttt{concat}(p_{\text{filter}}(\texttt{task}), y_r)$ \qquad \textcolor{gray}{$\triangleright$ $p_{\text{filter}}(\texttt{task})$: \texttt{task}-specific filtering prompt}
    \State $\texttt{response} \sim \mathcal{L}(p_r)$
    \If{$\texttt{response} = \text{
    YES
    }$}
        \State $D_{\mathrm{synth}} \leftarrow D_{\mathrm{synth}} \cup \{y_r\}$
    \EndIf
\EndFor
\label{alg:filter_end}
\State \textbf{Return:} $D_{\mathrm{synth}}$
\end{algorithmic}
\end{algorithm*}

\subsection{Keywords Extraction from Documents}
The keyword extraction step (line~\ref{alg:keyword}), which extracts $K$ keywords from each document $d_i$, is designed to reduce the sensitivity of the Gaussian mechanism applied during histogram generation. To achieve this, both the extraction prompt and the document are provided as input to the LLM. We employ the same prompt template across all datasets for this process. as shown below.

\begin{tcolorbox}[colback=gray!5!white, colframe=gray!30!black, title={Keywords Extraction Prompt}, breakable]
Extract \{$K$\} single words from the following document that represent key information specific to the content.

Document: \{$d_i$\}
\end{tcolorbox}

\subsection{Private Prediction}
\label{app:private_pred}
In private prediction, we prompt the LLM to rephrase each document $d_i$ in a subset (line~\ref{alg:pred_start}). The resulting token logits are then privately aggregated within the subset (line~\ref{alg:pred_agg}) to generate synthetic text (line~\ref{alg:pred_end}). 

\paragraph{Rephrasing Prompt.} We use the rephrasing prompt explicitly instructs the LLM to preserve the important information useful for downstream RAG tasks, as shown below. This prompt template is applied consistently across the entire dataset.

\begin{tcolorbox}[colback=gray!5!white, colframe=gray!30!black, title={Rephrasing Prompt}, breakable]
Rephrase the following document without altering the important information contained within it.

Document: \{$d_i$\}
\end{tcolorbox}

\paragraph{Clipping Method.} For the clipping method, we follow the approach of \citet{grislain25}, which emphasizes tokens with larger logit values and thereby reduces the influence of noise. The $t$-th element of the logit $l(t)$ is clipped as follows. First, we exponentiate $l(t)$ with a normalization factor to highlight large values:
\[
l^{\mathrm{exp}}(t) = \frac{e^{l(t)}}{\max_s e^{l(s)}}.
\]
Next, to reduce information loss from clipping within the range $[-c, c]$, we shift the center so that the maximum and minimum have equal magnitude:
\begin{align*}
    l^{\mathrm{cent}}(t) = l^{\mathrm{exp}}(t) 
    - \frac{\max_sl^{\mathrm{exp}}(s) + \min_sl^{\mathrm{exp}}(s)}{2}.
\end{align*}
Finally, we rescale $l^{\mathrm{cent}}(t)$ to lie within $[-c, c]$:
\begin{align*}
    \operatorname{clip}_c &\left(l\right)(t) =  l^{\mathrm{cent}}(t) \min\left(1, \frac{c}{||l^{\mathrm{cent}}||_{\infty}}\right).
\end{align*}

\subsection{Self-filtering}
Self-filtering filters synthetic texts by instructing the LLM to determine, based on task information, whether a synthetic text $y_r$ contains information relevant to the task (line~\ref{alg:filter_start}-\ref{alg:filter_end}). Because the filtering prompt varies across tasks, the prompts used for each dataset are listed below. Note that self-filtering is not applied to SearchQA, as it includes diverse questions and lacks a fixed task.

\begin{tcolorbox}[colback=gray!5!white, colframe=gray!30!black, title={Self-filtering Prompt on Medical Synth}, breakable]
Does the following document contain any specific diagnosis names, even if they are fictional? Answer only YES or NO.

Document: \{$y_r$\}

Answer:
\end{tcolorbox}

\begin{tcolorbox}[colback=gray!5!white, colframe=gray!30!black, title={Self-filtering Prompt on Movielens}, breakable]
Does the following document contain specific movie titles released in the 20th century? Answer only YES or NO.

Document: \{$y_r$\}

Answer:
\end{tcolorbox}

\section{Privacy Analysis}
\label{app:analysis}
In this section, we present the complete proof of Theorem~\ref{thrm}. We first introduce zero-concentrated differential privacy (zCDP) \cite{bun16}, a variant of DP that offers tighter composition bounds and more accurate privacy accounting, which we use in our analysis.

\begin{definition}[$\rho$-zCDP]
A randomized mechanism $\mathcal{M}$ satisfies $\rho$-concentrated differential privacy ($\rho$-zCDP) if for all $\alpha>1$
\[
D_{\alpha}(\mathcal{M}(D)||\mathcal{M}(D'))\leq\rho\alpha,
\]
where $D_{\alpha}(P||Q)$ is the R$\acute{e}$nyi divergence of order $\alpha$ between distributions $P$ and $Q$.  
\end{definition}

The widely used two DP algorithms defined in Section~\ref{sec:dp}: Gaussian mechanism and exponential mechanism both guarantee zCDP. 

\begin{lemma}[\cite{bun16}]
\label{gaussian}
Gaussian mechanism $\operatorname{GM}\colon\mathcal{D}\rightarrow\mathbb{R}^d$ of the form 
\[
\operatorname{GM}(D) = f(D)+\mathcal{N}(0, \sigma^2 I_d)
\]
satisfies $\rho$-zCDP for $\rho=\frac{(\Delta_2 f)^2}{2\sigma^2}$.
\end{lemma}

\begin{lemma}[\cite{cesar21}]
\label{exponential}
Exponential mechanism $\operatorname{EM}\colon\mathcal{D}\rightarrow\mathcal{Y}$ of the form 
\[
\operatorname{Pr}[\operatorname{EM}(D) = y] \propto \exp\left(\frac{\varepsilon \cdot u(D, y)}{2\Delta_{\infty} u}\right)
\]
satisfies $\rho$-zCDP for $\rho=\frac{1}{8}\varepsilon^2$.
\end{lemma}
\noindent Here, $\Delta_p f$ denotes the $L_p$-sensitivity of a function $f: \mathcal{D} \rightarrow \mathbb{R}^d$, defined as 
\[
\Delta_p f = \sup_{D, D'} \|f(D) - f(D')\|_p.
\] 

We note that $(\varepsilon,\delta)$-DP and $\rho$-zCDP can be converted into each other by the following lemma.

\begin{lemma}[Relationship between DP and zCDP \cite{bun16}]
\label{conversion}
Let $\mathcal{M}\colon\mathcal{D} \rightarrow \mathcal{Y}$ satisfy $\rho$-zCDP. Then $\mathcal{M}$ satisfies $(\varepsilon, \delta)$-DP for all $\delta > 0$ and
\[
\varepsilon = \rho + \sqrt{4 \rho \log(1/\delta)}.
\]
Thus, to achieve a given $(\varepsilon, \delta)$-DP guarantee, it suffices to satisfy $\rho$-zCDP with
\[
\rho = \left( \sqrt{\varepsilon + \log(1/\delta)} - \sqrt{\log(1/\delta)} \right)^2.
\]
\end{lemma}

As a final step, we introduce two composition theorems: the sequential composition theorem and the overlapping parallel composition theorem. The overlapping parallel composition is originally proposed by \citet{smith21}; in this paper, we adapt it to the zCDP framework and provide a proof.

\begin{lemma}[Sequential Composition \cite{bun16}]
\label{composition}
Let $\mathcal{M}\colon\mathcal{D}\rightarrow\mathcal{Y}$ and $\mathcal{M}'\colon\mathcal{D}\times\mathcal{Y}\rightarrow\mathcal{Z}$. Suppose $\mathcal{M}$ satisfies $\rho$-zCDP and $\mathcal{M}'$ satisfies $\rho'$-zCDP as a function of its first argument. Define $\mathcal{M}''\colon\mathcal{D}\rightarrow\mathcal{Z}$ by $\mathcal{M}''(D)=\mathcal{M}'(D,M(D))$. Then, it holds that
\begin{align}
    D_{\alpha}(\mathcal{M}''(D)||\mathcal{M}''(D'))&\leq D_{\alpha}(\mathcal{M}(D)||\mathcal{M}(D')) \notag \\
    +\sup_{y\in\mathcal{Y}}D_{\alpha}(\mathcal{M}'(D,y&)||\mathcal{M}'(D',y)),
\end{align}
and therefore, $\mathcal{M}''$ satisfies $(\rho+\rho')$-zCDP.
\end{lemma}

\begin{lemma}[Overlapping Parallel Composition]
\label{parallel}
Let $R$ and $L$ be positive integers. Let $P(d_i, r)$ be a proposition that depends only on $d_i \in D$ and $r\in[R]$, and is independent of other samples in $D$. For each $r\in[R]$, define a subset $C_r \subset D$ as
\begin{align*}
C_r &= \left\{ d_i \in D \mid P(d_i, r) = \mathrm{True} \right\}, \text{where}\\
\sum_{r=1}^{R} \mathbbm{1}&\left[P(d_i, r) = \mathrm{True} \right] \leq L \ \ \text{for any} \ \ d_i \in D.
\end{align*}
Let $\mathcal{M}\colon\mathcal{D}\rightarrow\mathcal{Y}$ be a mechanism that satisfies $\rho$-zCDP. If $\mathcal{M}'$ is the mechanism defined by
\[
\mathcal{M}'(D) = (\mathcal{M}(C_1),\dots,\mathcal{M}(C_R)),
\]
then $\mathcal{M}'$ satisfies $L\rho$-zCDP. 
\begin{proof}
    Let $D,D'$ be neighboring datasets. Without loss of generality, assume $D=D'\cup\{d_i\}$. Since the assignment of $d_i$ to each subset $C_r$ depends on only $d_i$ and $r$, it holds that $C_r=C_r'$ for $r$ such that $P(d_i, r) = \mathrm{False}$ and $C_r=C_r'\cup\{d_i\}$ for $r$ such that $P(d_i, r) = \mathrm{True}$. We have for all $\alpha > 1$
    \begin{align*}
        D_\alpha&(\mathcal{M}'(D)\|\mathcal{M}'(D')) \\
        &= \sum_{r=1}^RD_\alpha(\mathcal{M}(C_r)\|\mathcal{M}(C_r')) \\
        &= \sum_{r:P(d_i, r) = \mathrm{True}}D_\alpha(\mathcal{M}(C_r)\|\mathcal{M}(C_r')) \leq L\rho.
    \end{align*}
\end{proof}
\end{lemma}

We now prove Theorem~\ref{thrm}, establishing the privacy analysis of Algorithm~\ref{algorithm}.

\begin{proof}
    Our algorithm comprises two sequentially composed mechanisms: (a) histogram generation ($\mathcal{M}_{\text{hist}}$) and (b-d) keyword-based clustering followed by parallel operations on each cluster ($\mathcal{M}_{\text{clus}}$). Within $\mathcal{M}_{\text{clus}}$, each cluster undergoes a sequence of operations: (c) retrieval ($\mathcal{M}_{\text{retr}}$), (d) private prediction ($\mathcal{M}_{\text{pred}}$). Formally, the full algorithm $\mathcal{M}$ is defined as
    \begin{equation*}
        \mathcal{M}(D) = \mathcal{M}_{\text{clus}}(D,\mathcal{M}_{\text{hist}}(D)), 
    \end{equation*}
    and, given a histogram $h'$, $\mathcal{M}_{\text{clus}}$ is defined as
    \begin{align*}
        \mathcal{M}_{\text{clus}}(D,&h') = \\ (\mathcal{M}_{\text{pred}}&(\mathcal{M}_{\text{retr}}(C_1)),\dots,\mathcal{M}_{\text{pred}}(\mathcal{M}_{\text{retr}}(C_R))). \notag
    \end{align*}
    
    We first prove $\mathcal{M}_{\text{hist}}$, $\mathcal{M}_{\text{retr}}$, and $\mathcal{M}_{\text{pred}}$ satisfy $\rho_{\mathrm{hist}}$, $\rho_{\mathrm{retr}}$, $\rho_{\mathrm{pred}}$-zCDP, respectively. Since $\mathcal{M}_\text{hist}$ generates the histogram using the Gaussian mechanism with sensitivity $\sqrt{K}$, it satisfies $\rho_\text{hist}$-zCDP with $\rho_\text{hist} = \frac{K}{2\sigma_h^2}$ by Lemma~\ref{gaussian}. $\mathcal{M}_\text{retr}$ is a composition of the Gaussian mechanism and the exponential mechanism, both with sensitivity 1. Therefore, from Lemma~\ref{gaussian} and Lemma~\ref{exponential}, it satisfies $\rho_\text{retr}$-zCDP with $\rho_\text{retr} = \frac{1}{8}\varepsilon_{\theta_s}^2 + \frac{1}{2\sigma_{\mu}^2}$. 
    $\mathcal{M}_\text{pred}$ is a composition of $T$ applications of the exponential mechanism, so by Lemma~\ref{exponential}, it satisfies $\rho_\text{pred} = \frac{T}{2}\left(\frac{c}{\tau}\right)^2$-zCDP. 
    Hence, by Lemma~\ref{composition}, $\mathcal{M}_\text{clus}$ satisfies $(\rho_\text{ret} + \rho_\text{gen})$-zCDP.
    
    For the overall algorithm $\mathcal{M}$,Lemma~\ref{composition} gives
    \begin{align*}
        D_{\alpha}(\mathcal{M}(D)||\mathcal{M}(D'))\leq D_{\alpha}(\mathcal{M}_\text{hist}(D)||\mathcal{M}_\text{hist}(D')) \notag \\
        +\sup_{h'\in\mathbb{R}^{|\mathcal{V}|}}D_{\alpha}({\mathcal{M}_\text{clus}}(D,h')||\mathcal{M}_\text{clus}(D',h')).
    \end{align*}
    For any histogram $h'$, the assignment of $d_i \in D$ to a cluster $C_r$ depends only on $d_i$ and $r$. Thus, by Lemma~\ref{parallel},
    \begin{align*}
        &D_{\alpha}(\mathcal{M}(D)||\mathcal{M}(D'))\leq \rho_\text{hist} + L(\rho_\text{retr}+\rho_\text{pred}) \\
        &= \frac{K}{2\sigma_h^2} + L\left(\frac{1}{8}\varepsilon_{\theta_s}^2 + \frac{1}{2\sigma_{\mu}^2} + \frac{T}{2}\left(\frac{c}{\tau}\right)^2\right) = \rho.
    \end{align*}
    Therefore, the algorithm $\mathcal{M}$ satisfies $\rho$-zCDP, which can be converted to $(\varepsilon, \delta)$-DP by Lemma~\ref{conversion}.

\end{proof}

\section{\re{Experimental Details and Analyses}}
\label{app:experiment}
\subsection{Computational Resources}
All experiments in this study use 4 NVIDIA A100 GPUs with 40 GB memory each, running on a Linux-based server cluster equipped with
Intel Xeon Silver 4216 CPUs and 755 GB RAM. Reproducing the main results in Table \ref{tab:privacy_comparison} (5 runs of 5 methods across 3 datasets and 3 models) takes approximately 48 hours.

\subsection{Datasets}
\label{app:experiment_dataset}
In this paper, We use publicly available three datasets under their respective usage licenses: Medical Synth\footnote{Apache-2.0 license, \url{https://huggingface.co/datasets/sarus-tech/medical_dirichlet_phi3}}, Movielens\footnote{See the README for license details, \url{https://grouplens.org/datasets/movielens/}}, and SearchQA\footnote{BSD 3-Clause license, \url{https://github.com/nyu-dl/dl4ir-searchQA/tree/master}}
The details of them are summarized in Table~\ref{tab:datasets}.
Each dataset’s queries are divided into validation and test sets: 1,000 each for Medical Synth, 500 each for Movielens, and 102 each for SearchQA.
For Medical Synth and Movielens, patients or users not included in the query sets are used as the retrieval database for RAG. 
Table~\ref{tab:dataset_examples} shows examples of queries from each dataset and the corresponding top-1 retrieved document from the database.
The following describes the preprocessing details for Movielens.

\begin{table*}[t]
\centering
\setlength{\tabcolsep}{4pt} 
\renewcommand{\arraystretch}{1.25} 
\begin{tabularx}{\linewidth}{
    >{\raggedright\arraybackslash}p{1.5cm}  
    >{\centering\arraybackslash}p{2.0cm}    
    >{\centering\arraybackslash}p{2.0cm}    
    >{\centering\arraybackslash}p{2.5cm}    
    X                                       
}
\hline
\textbf{Dataset} & \textbf{\# Database} & \textbf{\# Query (Val/Test)} & \textbf{Answer Set} & \textbf{Task Description} \\
\hline

\textbf{Medical Synth} & 8,000 & 1,000 / 1,000 & 100 fictional disease names &
Given a patient's symptom description, retrieve similar doctor responses and output the correct fictitious disease name. Accuracy is evaluated by whether the correct disease name is included in the output. \\

\hline

\textbf{Movielens} & 4,083 & 500 / 500 & Top-30 frequent movie titles &
Recommend movies for a querying user based on their profile and generated preferences by retrieving similar users’ favorites. Accuracy is evaluated by whether the output includes any of the user’s top-10 favorite movies within the top-30 frequent titles. \\

\hline

\textbf{SearchQA} & 7,054 & 102 / 102 & Factual answers &
Answer Jeopardy! questions by retrieving search snippets. Accuracy is evaluated based on whether the model output includes the gold answer. \\
\hline
\end{tabularx}
\caption{Summary of datasets used for evaluation.}
\label{tab:datasets}
\end{table*}

\begin{table*}[t]
\centering
\setlength{\tabcolsep}{5pt}
\renewcommand{\arraystretch}{1.2}
\begin{tabularx}{\linewidth}{
    >{\raggedright\arraybackslash}p{2.0cm}  
    >{\raggedright\arraybackslash}X          
    >{\raggedright\arraybackslash}X          
}
\hline
\textbf{Dataset} & \textbf{Query Sample} & \textbf{Top-1 Retrieved Document} \\
\hline

\textbf{Medical Synth} &
I am Katarina Nordberg, I am dealing with severe itching specifically around my waistline, I also notice redness on my ears, and I find that my skin reacts unusually to cotton fabrics, exhibiting heightened sensitivity. What is my disease? &
Patient Fernando Lund is experiencing severe itching specifically around the waistline, redness in his ears, and heightened sensitivity to cotton fabrics. Based on these symptoms, the medical condition diagnosed is \textcolor{blue}{Flumplenoxis}. The recommended treatment for this condition is the administration of Doozy Drops. \\

\hline

\textbf{Movielens} &
This user is a 35-year-old male college student. He gravitates toward timeless, character-driven epics that blend action with adventure, sci-fi/fantasy, war, and crime, favoring heroic quests and moral complexity in richly realized worlds while also appreciating enduring family-friendly fantasy musicals. What movie is recommended for this user? Answer with movies released in the 20th century. &
Logan Butler is a 35-year-old male executive. He favors timeless, character-driven epics that blend action and adventure with rich world-building, moral complexity, and touches of wit and romance across sci-fi, crime drama, and fantastical adventure. In particular, he likes \textcolor{blue}{Star Wars: Episode IV - A New Hope (1977)}, \textcolor{blue}{The Godfather (1972)}, and The Princess Bride (1987). \\

\hline

\textbf{SearchQA} &
Question: The discovery of the Comstock Lode in 1859 attracted miners \& prospectors to this state &
In 1859, two young prospectors struck gold in the Sierra \textcolor{blue}{Nevada} lands. Henry Comstock discovered a vein of gold called a lode. The Comstock Lode attracted thousands of prospectors. Miners came across the United States, as well as from France, Germany, Ireland, Mexico, and China. One of every three miners was... \\
\hline
\end{tabularx}
\caption{Examples of queries and their top-1 retrieved documents in RAG. The correct answers contained in the documents are highlighted in blue.}
\label{tab:dataset_examples}
\end{table*}

\paragraph{Preprocessing of Movielens.} 
As the first step in converting MovieLens data into text, we use GPT-5 to generate textual descriptions of each user’s movie preferences from the user profile, the user’s 10 highest-rated films (restricted to the dataset’s 30 most frequently rated titles), and the genres of those films available in MovieLens. Because MovieLens does not include user names, we assign each user a GPT-5–generated pseudonym. We use the template below to generate these preferences.

\begin{tcolorbox}[colback=gray!5!white, colframe=gray!80!black, title={Movie Preference Generation Prompt}, breakable]
The following user is one of the users in the MovieLens dataset collected in 2000. Describe this user's movie preferences according to the following conditions:

1. Describe in one sentence that fully reflects their profile and the characteristics of the movies they like.

2. Do not include specific movie titles or the user’s profile information.

3. Begin with either "He" or "She".

4. Provide only the user’s preferences.
\vspace{1em}

User: \{name\} is a \{age\}-year-old \{gender\} \{occupation\}. He/She likes \{movie\_1\}, \{movie\_2\}, ...
\vspace{1em}

\{movie\_1\}

Genres: \{genre\_1\}, \{genre\_2\}, ...

...
\end{tcolorbox}
Using these generated preferences, we then create database documents with the template below. We use these documents as a RAG database to answer queries consisting of each user’s profile and generated preferences.

\begin{tcolorbox}[colback=gray!5!white, colframe=gray!80!black, title={Database Template for Movielens}, breakable]
\{name\} is a \{age\}-year-old \{gender\} \{occupation\}.

\{generated preference\}

In particular, he/she likes \{movie\_1\}, \{movie\_2\}, ...
\end{tcolorbox}


\subsection{Hyperparameter Search}
\label{app:_experiment_hyp}

Hyperparameters of each method are tuned using the validation queries.

\paragraph{Compared Methods.}
For DP-RAG, we mainly adopt the parameter values reported in the original paper \cite{grislain25}. The output token length is set to 70 for Medical Synth and MovieLens, and to 30 for SearchQA, as its answers are shorter. The top-p value, which controls the number of retrieved documents, is set to 0.02 for MedicalSynth (following the original paper) and 0.05 for MovieLens and SearchQA. For DP-Synth, the batch size is fixed at 100 for all datasets.

\paragraph{Proposed Method.}
The hyperparameters used for the main results of DP-SynRAG (Table~\ref{tab:privacy_comparison}) are summarized in Table~\ref{tab:hyperparams}. The other hyperparameters can be computed from those listed in the table. The parameters $\varepsilon_{\theta_s}$, $\rho_{\text{hist}}$, and $\rho_{\text{retr}}$ are set based on the total privacy budget. The parameters $K$ and $T$ are chosen according to the average token length per record in the dataset and kept constant across all experiments. The parameter $R$ determines the number of words extracted from the noisy histogram. To minimize the probability of extracting words with original zero counts, $R$ is set at the position where the word frequency approximately corresponds to $3\sigma_h$ when the words are sorted by frequency. For Medical Synth and MovieLens, this corresponds to $R=500$, and for SearchQA, $R=1000$.
The parameters $L$ and $k$ are tuned using the validation queries. To assess the sensitivity of DP-SynRAG to its hyperparameters $K$, $R$, $L$, and $k$, Tables~\ref{tab:sensitivity_K_std}-\ref{tab:sensitivity_k_std} report the average accuracy on the test queries as each hyperparameter varies. \re{Our sensitivity analysis shows that performance deteriorates only under extreme values (e.g., $L=1$, corresponding to hard clustering). This behavior is consistent across datasets, which suggests that it is possible to identify a set of safe hyperparameter ranges that work robustly in practice, without extensive dataset-specific tuning.}

\begin{table*}[t]
\centering
\begin{tabular}{lccccccccc}
\hline
\textbf{Dataset} & $\tau$ & $K$ & $R$ & $L$ & $k$ & $T$ & $\varepsilon_{\theta_s}$ & $\rho_{\text{hist}}$ & $\rho_{\text{retr}}$ \\
\hline
Medical Synth & 1.0 & 10 & 500 & 5 & 80 & 70 & 0.4 & 0.1 & 0.009 \\
Movielens     & 1.0 & 10 & 500 & 5 & 100 & 70 & 0.4 & 0.1 & 0.009 \\
SearchQA      & 1.0 & 10 & 1000 & 5 & 100 & 70 & 0.4 & 0.1 & 0.009 \\
\hline
\end{tabular}
\caption{Hyperparameters of DP-SynRAG for the main results presented in Table~\ref{tab:privacy_comparison}. $\tau$: temperature parameter for private prediction; $K$: number of keywords extracted from each document; $R$: number of clusters; $L$: number of overlapping documents across clusters; $k$: number of retrieved documents; $T$: token length of synthetic texts; $\varepsilon_{\theta_s}$: privacy parameter of threshold selection; $\rho_{\text{hist}}$: zCDP parameter of the histogram generation step; $\rho_{\text{retr}}$: zCDP parameter of the retrieval step.}
\label{tab:hyperparams}
\end{table*}

\begin{table*}[htbp]
\centering
\begin{tabular}{lccc}
\hline
\textbf{Dataset} & $K=5$ & $K=10$ & $K=20$ \\
\hline
Medical Synth & 67.18\,\diff{black}{2.29} & 67.26\,\diff{black}{2.22} & 67.84\,\diff{black}{1.68} \\
Movielens & 44.48\,\diff{black}{1.71} & 42.56\,\diff{black}{1.97} & 44.16\,\diff{black}{1.79} \\
SearchQA & 91.57\,\diff{black}{2.26} & 89.61\,\diff{black}{3.22} & 92.16\,\diff{black}{1.83} \\
\hline
\end{tabular}
\caption{Sensitivity analysis of DP-SynRAG with respect to the number of keywords extracted from each document $K$. We set $\varepsilon_{\text{total}}=10$. All other hyperparameters except $K$ are set to the values listed in Table~\ref{tab:hyperparams}.}
\label{tab:sensitivity_K_std}
\end{table*}

\begin{table*}[htbp]
\centering
\begin{tabular}{lccccc}
\hline
\textbf{Dataset} & $R=100$ & $R=300$ & $R=500$ & $R=700$ & $R=1000$ \\
\hline
Medical Synth & 52.84\,\diff{black}{3.47} & 67.56\,\diff{black}{1.22} & 67.26\,\diff{black}{2.22} & 66.68\,\diff{black}{2.91} & 66.36\,\diff{black}{2.00} \\
Movielens & 44.16\,\diff{black}{2.18} & 44.60\,\diff{black}{1.74} & 42.56\,\diff{black}{1.97} & 40.56\,\diff{black}{1.31} & 44.32\,\diff{black}{2.42} \\
SearchQA & 72.94\,\diff{black}{1.12} & 88.24\,\diff{black}{4.27} & 90.78\,\diff{black}{1.78} & 90.00\,\diff{black}{1.89} & 89.61\,\diff{black}{3.22} \\
\hline
\end{tabular}
\caption{Sensitivity analysis of DP-SynRAG with respect to the number of clusters $R$. We set $\varepsilon_{\text{total}}=10$. All other hyperparameters except $R$ are set to the values listed in Table~\ref{tab:hyperparams}.}
\label{tab:sensitivity_R_std}
\end{table*}

\begin{table*}[htbp]
\centering
\begin{tabular}{lccccc}
\hline
\textbf{Dataset} & $L=1$ & $L=3$ & $L=5$ & $L=7$ & $L=10$ \\
\hline
Medical Synth & 42.52\,\diff{black}{4.84} & 62.82\,\diff{black}{2.97} & 67.26\,\diff{black}{2.22} & 67.86\,\diff{black}{1.88} & 66.14\,\diff{black}{2.25} \\
Movielens & 38.36\,\diff{black}{2.54} & 40.44\,\diff{black}{3.10} & 42.56\,\diff{black}{1.97} & 44.96\,\diff{black}{2.10} & 43.12\,\diff{black}{3.42} \\
SearchQA & 76.67\,\diff{black}{2.54} & 85.62\,\diff{black}{1.50} & 89.61\,\diff{black}{3.22} & 88.56\,\diff{black}{2.04} & 85.29\,\diff{black}{1.70} \\
\hline
\end{tabular}
\caption{Sensitivity analysis of DP-SynRAG with respect to the number of overlapping documents across clusters $L$. We set $\varepsilon_{\text{total}}=10$. All other hyperparameters except $L$ are set to the values listed in Table~\ref{tab:hyperparams}.}
\label{tab:sensitivity_L_std}
\end{table*}

\begin{table*}[htbp]
\centering
\begin{tabular}{lccccc}
\hline
\textbf{Dataset} & $k=40$ & $k=60$ & $k=80$ & $k=100$ & $k=120$ \\
\hline
Medical Synth & 57.84\,\diff{black}{3.00} & 65.68\,\diff{black}{1.58} & 67.26\,\diff{black}{2.22} & 68.16\,\diff{black}{2.38} & 67.42\,\diff{black}{1.88} \\
Movielens & 40.32\,\diff{black}{1.30} & 41.84\,\diff{black}{1.86} & 44.20\,\diff{black}{1.17} & 42.56\,\diff{black}{1.97} & 43.48\,\diff{black}{2.59} \\
SearchQA & 87.25\,\diff{black}{3.47} & 90.78\,\diff{black}{0.88} & 89.61\,\diff{black}{2.03} & 89.61\,\diff{black}{3.22} & 90.59\,\diff{black}{1.91} \\
\hline
\end{tabular}
\caption{Sensitivity analysis of DP-SynRAG with respect to the number of retrieved documents $k$. We set $\varepsilon_{\text{total}}=10$. All other hyperparameters except $k$ are set to the values listed in Table~\ref{tab:hyperparams}.}
\label{tab:sensitivity_k_std}
\end{table*}

\subsection{Direct Evaluation of Synthetic Text Quality}
\label{app:synthetic_quality}

\re{To complement our downstream RAG evaluation, we additionally evaluate the synthetic texts directly using perplexity. We compare four corpora: (1) the original private database, (2) DP-Synth, (3) the full DP-SynRAG synthetic database, and (4) the subset of DP-SynRAG texts that contain the ground-truth answer words. Lower perplexity should be interpreted only as a complementary signal in our setting, because generic but uninformative text can also attain low perplexity.}

\re{Table~\ref{tab:synthetic_quality} shows that the full DP-SynRAG database can have relatively high perplexity, reflecting the presence of some low-quality entries generated from small or noisy subsets. However, when we focus on task-relevant DP-SynRAG texts containing the ground-truth answer words, perplexity decreases substantially across datasets and models. In practice, RAG uses only the data that is highly relevant to the query as context. Moreover, filtering based on perplexity or performing LLM refinement passes could further improve quality.}

\begin{table*}[t]
\centering
\small
\setlength{\tabcolsep}{4pt}
\begin{tabular}{lccc|ccc|ccc}
\toprule
& \multicolumn{3}{c|}{Medical Synth} & \multicolumn{3}{c|}{Movielens} & \multicolumn{3}{c}{SearchQA} \\
Method & Phi-4 & Gemma-2 & Llama-3.1 & Phi-4 & Gemma-2 & Llama-3.1 & Phi-4 & Gemma-2 & Llama-3.1 \\
\midrule
Original & 31.16 & 72.81 & 29.97 & 19.82 & 20.47 & 19.76 & 51.13 & 48.80 & 31.60 \\
DP-Synth & 10.64 & 13.15 & 11.35 & 5.94 & 7.15 & 7.67 & 4.15 & 6.26 & 7.75 \\
DP-SynRAG & 70.80 & 153.49 & 23.94 & 17.01 & 44.36 & 19.36 & 71.62 & 106.39 & 117.54 \\
DP-SynRAG (GT) & 45.89 & 65.44 & 21.70 & 30.58 & 23.87 & 14.87 & 70.47 & 106.85 & 91.10 \\
\bottomrule
\end{tabular}
\caption{\re{Perplexity of the original private database and synthetic database. Lower is better. ``DP-SynRAG (GT)'' denotes the subset of DP-SynRAG texts that contain the ground-truth answer words.}}
\label{tab:synthetic_quality}
\end{table*}

\re{
\subsection{Runtime and Database Refresh}
\label{app:runtime_refresh}

In practical RAG deployments, the underlying database may be updated over time, making the cost of synthetic database construction important in addition to final task accuracy. DP-SynRAG consists of two main stages: (i) DP clustering and retrieval, and (ii) synthetic text generation.

\paragraph{Computational Overhead.}
Let $N$ denote the number of private documents. The clustering stage consists of: (1) keyword extraction, (2) histogram construction and top-$R$ keyword selection, (3) keyword-based soft cluster assignment, and (4) embedding-based refinement. Among these, keyword extraction and embedding-based refinement dominate the cost in practice, each requiring $O(N)$ LLM forward passes or embedding computations. The synthetic generation stage processes documents assigned to each cluster. Since each document belongs to at most $L$ clusters, the total number of LLM calls is bounded by $O(NL)$.

On Medical Synth ($N=8{,}000$), using Llama-3.1-8B, a single NVIDIA A100 40GB GPU, and a 64-core Intel Xeon Silver 4216 CPU, clustering takes 508 seconds and synthetic generation takes 1,944 seconds. Thus, LLM inference dominates the total cost.

\paragraph{Frequently Updated Databases.}
When the private database is updated, a full rerun is not always necessary. New documents only require keyword extraction and embedding computation before being assigned to existing clusters, and synthetic generation can then be rerun only for the affected subsets. Therefore, although DP-SynRAG introduces a nontrivial preprocessing cost, it avoids model retraining and can be refreshed more efficiently than private fine-tuning approaches in frequently updated settings.

At the same time, database updates remain a practical deployment challenge for any DP synthetic generation pipeline. Our contribution mainly shifts DP enforcement from query time to data-generation time; improving incremental refresh mechanisms is an important direction for future work.
}

\re{
\subsection{Additional Comparison with Query-time DP Baselines}
\label{app:query_time_baselines}

To complement the main results, we additionally compare DP-SynRAG with another query-time DP baseline, DP-SparseVote \cite{koga25}, which privatizes LLM outputs directly at inference time. In this comparison, DP-SynRAG uses a total privacy budget of $\varepsilon_{\text{total}}=10$, whereas DP-RAG and DP-SparseVote use a per-query privacy budget of $\varepsilon_{\text{query}}=10$. 

\begin{table*}[t]
\centering
\small
\setlength{\tabcolsep}{4pt}
\begin{tabular}{lccc|ccc|ccc}
\toprule
& \multicolumn{3}{c|}{Medical Synth} & \multicolumn{3}{c|}{Movielens} & \multicolumn{3}{c}{SearchQA} \\
Method & Phi-4 & Gemma-2 & Llama-3.1 & Phi-4 & Gemma-2 & Llama-3.1 & Phi-4 & Gemma-2 & Llama-3.1 \\
\midrule
DP-RAG        & 59.92 & 67.06 & 48.94 & 34.72 & 40.48 & 56.80 & 85.10 & 83.14 & 84.90 \\
DP-SparseVote & 48.59 & 32.68 & 24.62 & 34.29 & 38.32 & 55.76 & 94.51 & 84.51 & 94.51 \\
DP-SynRAG     & 67.26 & 67.06 & 61.26 & 42.56 & 41.08 & 54.12 & 89.61 & 85.10 & 91.18 \\
\bottomrule
\end{tabular}
\caption{\re{Additional comparison of accuracy (\%) with query-time DP baselines. DP-SynRAG uses a fixed total privacy budget of $\varepsilon_{\text{total}}=10$, whereas DP-RAG and DP-SparseVote use a per-query privacy budget of $\varepsilon_{\text{query}}=10$.}}
\label{tab:query_time_baselines}
\end{table*}

DP-SynRAG outperforms the additional query-time baselines on Medical Synth and Movielens, while DP-SparseVote shows better performance on SearchQA. These results reinforce our main claim that one-time synthetic database generation is especially effective in multi-query RAG settings, where query-time privatization repeatedly consumes privacy budget.
}

\section{Examples of Synthetic Texts}
\label{app:examples}
Table~\ref{tab:synth_examples} presents synthetic data samples generated by our proposed method. We include both good examples, which preserve essential information, and bad examples, which lose key information and are of low quality as text due to being generated from a small subset. In the good examples, sensitive information such as names is removed or replaced with LLM-generated pseudonyms, thereby protecting privacy.

\begin{table*}[t]
\centering
\setlength{\tabcolsep}{5pt}
\renewcommand{\arraystretch}{1.2}
\begin{tabularx}{\linewidth}{
    >{\raggedright\arraybackslash}p{2.0cm}  
    >{\raggedright\arraybackslash}X          
    >{\raggedright\arraybackslash}X          
}
\hline
\textbf{Dataset} & \textbf{Good Synthetic Text} & \textbf{Bad Synthetic Text} \\
\hline

\textbf{Medical Synth} &
\textcolor{ForestGreen}{Patient K, displaying symptoms of sudden limb weakness, uncontrolled gas release, and a peculiar tingling sensation in the left nasal passage, has been diagnosed with Flibberflamia Frigibulitis. To effectively manage this condition, a treatment plan tailored to address the specific needs of Flibberflamia Frigibulitis} &
\textcolor{red}{A case file lists certain anomalies L'Andre Duche whose assumed french inspired nickname appears incorrect was mistaken it has another " name given in it it indicates he suffering in issues of, haying filds for distance. as result confusion problems his way see think, problems his of breath have an haze,. Following a set list by certain of symptoms} \\

\hline

\textbf{Movielens} &
\textcolor{ForestGreen}{Zachary \"Scarlett \"Lee is a 25-year-old male. He has a strong affinity for dark, complex, and thought-provoking films that often blend elements of drama, crime, and the supernatural. His favorite movies include American Beauty (1999), and The Usual Suspects (1995).} &
\textcolor{red}{18-month college female participant Val Addabelle isn' covered ( but  information does show) sales Associate like   to paint as it reminds    me and the rest with in, \" action thrill rides   she most finds appealing The genre   for suspense \" movie that was shown at school by,The classic in year    she in was able a bit for} \\

\hline

\textbf{SearchQA} &
\textcolor{ForestGreen}{On this initial mention of renowned developer Mikhail Kalash transformer of the iconic device, the first AK-47 assault rifle was created in him Russia, The AK-47 was first introduced to Russian forces in 1949.} &
\textcolor{red}{Following specific guidelines a toy sweetening alternative, referred to as conjunctions - artificially and naturally used to reduce the meaning without an interruption of the overall message - compounds are used in sugarcoaster and they produce the same result when two or no items - (a) are compared to (x)= another item; 
2. Two substances with the combination} \\
\hline
\end{tabularx}
\caption{Examples of synthetic texts generated by DP-SynRAG for three datasets. Good examples (green) preserve essential information for downstream RAG tasks. Bad examples (red) lose key information and are of low quality as text.}
\label{tab:synth_examples}
\end{table*}

\end{document}